\documentclass[aps,twocolumn]{revtex4}
\def\Tr{{\rm Tr}\, }
\newcommand{\be}{\begin{eqnarray}}
\newcommand{\ee}{\end{eqnarray}}

\newcommand{\lb}{\label}
\def\>{\rangle}
\def\<{\langle}

\def\tr{\hbox{Tr}}

\def\be{\begin{eqnarray}}
\def\ee{\end{eqnarray}}
\def\lb{\label}
\begin{document}
\title{Entanglement Entropy in Non-Relativistic Field Theories}
\author{
Sergey N. Solodukhin
} \affiliation{{\it Laboratoire de Math\'ematiques et Physique
Th\'eorique CNRS-UMR 6083,  Universit\'e de Tours, Parc de
Grandmont, 37200 Tours, France \\}}
%\maketitle

\begin{abstract}
\noindent { We calculate entanglement entropy in a
non-relativistic field theory  described by the Schr\"{o}dinger
operator. We demonstrate that the entropy is characterized by i)
the area law and ii) UV divergences  that are identical to those
in the relativistic field theory. These observations are further
supported by a holographic consideration. We use the
non-relativistic symmetry and completely specify entanglement
entropy in large class of non-relativistic theories described by
the field operators polynomial in derivatives.  The entropy of
interacting fields is analyzed in some detail.
 We suggest that the area law of the
entropy can be tested in experiments with condensed matter systems
such as liquid helium.

}
\end{abstract}
\vskip 2.pc
\maketitle

\noindent {\it Introduction.} Entanglement entropy  of a field
system is defined by tracing over modes  that reside  inside a
chosen surface $\Sigma$ \cite{BS}. If the total system is
characterized by a pure quantum state the subsystem inside the
surface is described by a density matrix $\rho$ with the von
Neumann entropy $S=-\Tr \rho\ln\rho$ known as {\it entanglement
entropy} (see recent reviews \cite{Casini:2009sr},
\cite{Nishioka:2009un} and references therein). This entropy is
non-vanishing provided there are short-distance correlations in
the system. The presence of such correlations has two
consequences: a) entanglement entropy is determined by geometry of
the surface $\Sigma$, to leading order by the area of $\Sigma$ and
in higher orders by intrinsic and extrinsic geometry of $\Sigma$,
see \cite{Solodukhin:2008dh}; and b) the entropy is UV divergent.
 In d space-time dimensions, for a field system described by a relativistic
wave equation, one finds that  \be S\sim N\  {Area ( {\Sigma})
\over \epsilon^{d-2}}~~. \label{S} \ee The exact coefficient of
proportionality in (\ref{S}) depends on the regularization scheme,
$N$ is the number of field species. Entanglement entropy has been
actively studied in the literature. The main focus has been made
on the relativistic field systems and especially the conformal
field theories. In the latter case a holographic description of
entanglement entropy has been proposed that suggests an
alternative way to compute the entropy in the strongly coupled
conformal field  theory dual to supergravity on anti-de Sitter
space-time \cite{Nishioka:2009un}.

The definition of entanglement entropy is applicable to any system
that has a quantum mechanical description and does not necessarily
tie to the Lorentz symmetry. A typical example considered in the
literature is a finite system of coupled non-relativistic
oscillators \cite{BS}. The Hamiltonian of this system is quadratic
in momenta. The continuous limit then reproduces a relativistic
field described by the Klein-Gordon  operator which is quadratic
in time derivative. The primary purpose of the present note is to
compute the entanglement entropy for a field system described by
the Schr\"{o}dinger field operator linear in time derivative. The
field system in this case can not be presented as a collection of
harmonic oscillators. Nevertheless, as we show below,  the
structure of the entropy in this case is essentially identical to
the one in a relativistic field theory.

The AdS/CFT correspondence can be generalized to non-relativistic
conformal field theories  \cite{Son:2008ye} (for an
earlier work see \cite{Duval:1990hj}), see also
\cite{Kachru:2008yh}. These theories describe the ``fermions at
unitarity''. In a wider context the non-relativistic theories
effectively describe many realistic systems well studied
experimentally and theoretically in condensed matter,  liquid
helium He$^4$ is the standard example. Our results thus are
applicable to these systems too.

The field models with the broken Lorentz symmetry are sometimes
considered as UV completions of otherwise non-renormalizable
relativistic effective theories, e.g. the General Relativity.
We anticipate that our results can be extended to
those theories.

\bigskip

\noindent{\it The replica method.} Before proceeding we remind the
technical method very useful for calculation of entanglement
entropy. This method is known as {\it the replica method}, see
ref.\cite{Callan:1994py}. One first observes that $-\Tr \rho \ln
\rho=(\alpha\partial_\alpha-1)\Tr \rho^\alpha|_{\alpha=1}$. The
next observation is that  the density matrix obtained by tracing
over modes inside the surface $\Sigma$ is
$\tr\rho^\alpha=W[\alpha]$, where $W[\alpha]=-\ln Z(\alpha)$ and
$Z(\alpha)$ is the partition function of the field system in
question considered on Euclidean space with a conical singularity at the
surface $\Sigma$. Thus one has that \be
S=(\alpha\partial_\alpha-1)W(\alpha)|_{\alpha=1}~~. \lb{SS} \ee
 One chooses the local
coordinate system $\{X^\mu=(\tau,x_i)\},$ where $\tau$ is
Euclidean time, such that the surface $\Sigma$ is defined by
conditions $\tau=0,x_1=0$ and $(x_2,..,x_d)$ are the coordinates
on $\Sigma$. In the subspace $(\tau,x_1)$ it is convenient to
choose the polar coordinate system $\tau=r\sin(\phi)$ and
$x_1=r\cos(\phi)$ where angular coordinate $\phi$ changes in the
limits $0\leq \phi < 2\pi$. The conical space in question is then
defined by making the coordinate $\phi$ periodic with the period
$2\pi\alpha$, where $(1-\alpha)$ is very small.

In order to calculate the effective action $W(\alpha)$ we use the
heat kernel method. For  manifolds with conical singularities this
method was earlier developed in great detail in \cite{coneQFT}.
Consider a quantum bosonic field  described by a field operator
$\cal D$ so that $Z=\det^{-1/2}{\cal D}$. Then the effective
action is defined as \be W=-{1\over 2}\int_{\epsilon^2}^\infty
{ds\over s}\Tr K\lb{W}~~, \ee where $\epsilon$ is an UV cut-off,
is expressed by means of the trace of the heat kernel
$K(X,X',s)=\langle X|e^{-s{\cal D}}|X'\rangle$ satisfying the heat
kernel equation \be &&(\partial_s+{\cal
D})K(X,X',s)=0~,\nonumber \\
&&K(X,X',s=0)=\delta(X,X') ~~.\lb{K} \ee In the Lorentz invariant
case the heat kernel $K(\phi,\phi',s)$ (where we skip the
coordinates other than angle $\phi$) on regular flat space depends
on the difference $(\phi-\phi')$. The heat kernel $K_\alpha
(\phi,\phi,s)$ on space with a conical singularity is then
constructed from this quantity by applying the Sommerfeld formula
\cite{Sommerfeld}
\be &&K_\alpha(\phi,\phi',s)=K(\phi-\phi',s)\nonumber \\
&&+{\imath \over 4\pi\alpha}\int_\Gamma \cot {w\over 2\alpha}
K(\phi-\phi'+w,s)dw~~. \lb{Sommerfeld} \ee The contour $\Gamma$
consists of two vertical lines, going from $(-\pi+\imath \infty )$
to $(-\pi-\imath \infty )$ and from $(\pi-\imath \infty )$ to
$(\pi-+\imath \infty )$ and intersecting the real axis between the
poles of the $\cot {w\over 2\alpha}$: $-2\pi\alpha$, $0$ and $0,$
$+2\pi\alpha$ respectively. For $\alpha=1$ the integrand in
(\ref{Sommerfeld}) is a $2\pi$-periodic function and the
contributions of these two vertical lines cancel each other. Thus,
for a small angle deficit the contribution of the integral in
(\ref{Sommerfeld}) is proportional to $(1-\alpha)$.

\bigskip

\noindent{\it The Laplace operator.} In $d$ space-time dimensions,
for a relativistic theory described by the massless Klein-Gordon
operator or, in Euclidean signature, by the  Laplace operator
$\nabla^2=\partial^2_\tau+\sum_{i=1}^{d-1}\partial^2_{i}$, where
$\tau$ is Euclidean time, the heat kernel is known exactly \be
K(\tau,\tau',x,x',s)={1\over (4\pi s)^{d/2}}e^{-{1\over
4s}[(\tau-\tau')^2+\sum_i (x_i-x'_i)^2]} \lb{Kd} \ee We take
$(d-2)$-surface $\Sigma$ to be the plane defined by equations
$x_1=0, \ \tau=0$ so that $(x_2,x_3,..,x_d)$ are coordinates on
$\Sigma$.  In the polar coordinate system $\tau=r\sin\phi$ and
$x_1=r\cos\phi$ we have for two points $(r,\phi)$ and $(r,\phi')$
that $(\tau-\tau')^2+(x_1-x'_1)^2=4r^2\sin^2({\phi-\phi'\over
2})$. The trace is defined as $\Tr K_\alpha=\int
d^{d-2}x_i\int_0^\infty dr\ r\ \int_0^{2\pi\alpha}d\phi
K_\alpha(\phi=\phi',r'=r,x_i=x'_i,s)$. For the contour integral
$\Gamma$ we find that \cite{coneQFT} \be C_2(\alpha)\equiv{\imath
\over 8\pi\alpha}\int_\Gamma \cot{w\over 2\alpha}\ {dw\over
\sin^2{w\over 2}}={1\over 6\alpha^2}(1-\alpha^2)~~. \lb{C} \ee
Thus one obtains for the trace of the heat kernel (\cite{coneQFT})
\be \Tr K_\alpha={1\over (4\pi s)^{d/2}}\left(\alpha V+s \
2\pi\alpha C_2(\alpha)A(\Sigma)\right)~~, \lb{TrK} \ee where
$V=\int d\tau d^{d-1}x$ is the volume of spacetime and
$A(\Sigma)=\int d^{d-2}x$ is the area of surface $\Sigma$, and
 \be
S_{(d)}={2\over 3(d-2)}{\pi \over(4\pi)^{d/2}}{A(\Sigma)\over
\epsilon^{d-2}}~~ \lb{ent}\ee for entanglement entropy in $d$
space-time dimensions

\bigskip

\noindent{\it The Euclidean Schr\"{o}dinger operator.} We consider
a non-interacting system described by a non-relativistic field
operator ${\cal D}=-2im\partial_t-\sum_{i=1}^{d-1}\partial^2_i$.
Written in this form the Schr\"{o}dinger operator $\cal D$ can be
viewed as a reduction of the massless Klein-Gordon operator in
$d+1$ dimensions $2\partial_\xi\partial_t-\partial^2_i$ after
projecting onto the space of fixed momentum $\partial_\xi=-im$ as
described in \cite{Son:2008ye}, \cite{Duval:1990hj}. The Euclidean
Schr\"{o}dinger positive operator is further defined  as \be {\cal
D}_E=2m\sqrt{-\partial_\tau^2} -\partial_i^2~,\lb{D}\ee which is
formally obtained by   a double analytic continuation
$t\rightarrow i \tau$, $m\rightarrow -i m$.
 The heat kernel of this operator is
a product of the heat kernels of two operators
$$e^{-s{\cal
D}_E}=e^{-s2m\sqrt{-\partial^2_\tau}}\cdot
e^{-s(-\partial_i^2)}~~.
$$
The heat kernel of the $(d-1)$-dimensional Laplace operator
$\partial^2_i$ is given by expression similar to (\ref{Kd}).
As for the heat kernel of the first order operator $2\imath m
\partial_\tau$ it is convenient first to Laplace transform \cite{GR}
it to the heat kernel of the second order operator
$-\partial^2_\tau$, \be e^{-s2m\sqrt{-
\partial^2_\tau}}=\int_0^\infty d\sigma {m\over \sqrt{\pi}}{s\over
\sigma^{3/2}}e^{-{s^2 m^2\over \sigma}}\ e^{-\sigma
(-\partial_\tau^2)} \lb{Laplace} \ee and then use expression of
the type  (\ref{Kd}) for the heat kernel of operator
$-\partial^2_\tau$. Switching to the polar coordinates $(r,\phi)$,
$\tau=r\sin\phi$, $x_1=r\cos\phi$, we find for the heat kernel
function of two points characterized by different angular
coodinates $\phi$ and $\phi'$ $(r'=r,x'_i=x_i)$, \be &&\langle
\phi'|e^{-t{\cal D}_E}|\phi \rangle={2ms^2\over (4\pi s)^{d+1\over
2}}\int_0^\infty {d\sigma\over \sigma^2}e^{-(sm)^2/
\sigma}\nonumber
\\
&&e^{-r^2\sin^2{\chi\over 2}({1\over \sigma}\cos^2{\psi\over
2}+{1\over s} \sin^2{\psi\over 2})}~~, \lb{mess} \ee where we
introduced $\chi=(\phi-\phi')$ and $\psi=(\phi'+\phi)$.

The kernel now depends on two angular variables $\psi$ and $\chi$.
This is a consequence of the absence of the Lorentz symmetry.
Constructing the heat kernel on space with a conical singularity
we have to make the heat kernel to be periodic function of both
angles, $\chi$ and $\psi$,   with the period $2\pi\alpha$. We are
however interested in the result to leading order in $(1-\alpha)$.
The Sommerfeld formula  applied to the angular coordinate $\chi$
already gives the term proportional to $(1-\alpha)$. Thus, it is
sufficient to apply the formula (\ref{Sommerfeld}) only to the
angular coordinate $\chi$ and consider the other angular variable
$\psi$ as periodic with the period $2\pi$. Calculating the trace
one thus first replaces $\chi$ by $ \chi+w$, takes $\phi'=\phi$,
integrates over $\phi$ from $0$ to $2\pi$ and then applies formula
(\ref{Sommerfeld}) to $w$.
 Using the
integral
$$
\int_0^{2\pi}d\phi\int_0^\infty dr r e^{-r^2\sin^2{w\over
2}({1\over \sigma}\cos^2{\phi}+{1\over s}
\sin^2{\phi})}={\pi\sqrt{s\sigma}\over \sin^2{w\over 2}}
$$
and $\int_0^\infty d\sigma
\sigma^{-\nu}e^{-(sm)^2/\sigma}=(sm)^{2(1-\nu)}\Gamma (\nu-1)$ and
formula (\ref{C}) we arrive at the expression for the trace of the
heat kernel \be \Tr e^{-s{\cal D}_E}={2m^{-1}\alpha V\over (4\pi
s)^{d+1\over 2}}+{s\over (4\pi s)^{d\over 2}}2\pi
C_2(\alpha)A(\Sigma)~, \lb{TrKD} \ee valid up to terms of order
$(1-\alpha)^2$. We see that the volume part of the trace of the
heat kernel has the dependence on the proper time $s$ typical for
the heat kernel of the Laplace operator in $(d+1)$ dimensions.
This seems to be consistent with the way we obtained the
Schr\"{o}dinger operator as a reduction from the Klein-Gordon
operator in $(d+1)$ dimensions.

On the other hand, rather surprisingly, the surface term in
(\ref{TrKD})   is precisely identical to the surface term in the
heat kernel of the $d$-dimensional Laplace operator (\ref{TrK}).
The entanglement entropy in  a non-relativistic theory
 in any
dimension $d$ is thus given by the same formula (\ref{ent}) as in
the case of a real scalar field described by the  Lorentz
invariant Laplace operator. In four space-time dimensions we hence
obtain \be S_{(d=4)}={A(\Sigma)\over 48\pi\epsilon^2}~~. \lb{Sent}
\ee In usual signature the non-relativistic field is naturally
described by a complex function. For a complex field everything we
have just calculated then should be multiplied by $2$.

 The fact
that the entropy does not depend on the mass parameter $m$ has a
simple explanation.  The operator ${\cal D}_E$ is invariant under
the rescaling $\tau\rightarrow \beta \tau$, $m\rightarrow \beta \
m$. Note that both terms in (\ref{TrKD}) are invariant under this
rescaling. Since the entropy is not supposed to be a function of
$\tau$ (or $t$ in the usual signature) it  is not  a function
of $m$ either.

\bigskip

\noindent{\it A holographic consideration.} Recently there have
been proposed some dual gravity descriptions of non-relativistic
systems. In this note we use the approach suggested in
\cite{Kachru:2008yh}. As an example we consider $5$-dimensional
geometry described by metric \be ds^2=L^2({dr^2\over r^2}+{1\over
r^2} \sum_{i=1}^3 dx_i^2-{dt^2\over r^{2z}})~~, \lb{metric} \ee
which is invariant under the scale transformations \be
t\rightarrow \lambda^z t,\ x_i\rightarrow \lambda x_i, \
r\rightarrow \lambda r~~. \lb{trans} \ee For $z=2$ these are also
the symmetries of the Schr\"{o}dinger operator $\cal D$. Thus, one
expects that (for $z=2$) the metric (\ref{metric}) gives us a
gravity description of the non-relativistic system. For $z=1$
metric (\ref{metric}) is the anti-de Sitter metric which has
larger symmetry $SO(4,2)$. In this case the AdS/CFT correspondence
is at work with a precise dictionary of translation between the
quantum field theory and the gravity descriptions.

In the AdS/CFT correspondence it is known that entanglement
entropy has a holographic gravity description (for a recent review
see \cite{Nishioka:2009un}). The entropy is defined in the quantum
theory living on the boundary of the $5$-dimensional space-time
(\ref{metric}). Since the geometric quantities are usually
divergent when extended till the boundary one considers a
regularized boundary at $r=\epsilon$. According to the AdS/CFT
dictionary,  $\epsilon$ plays the role of the UV regulator on the
field theory side. For a given $2d$-surface $\Sigma$ on the
regularized boundary consider  a minimal $3d$-surface $\gamma$
such that it lies entirely on the slice $t=const$ and
$\partial\gamma=\Sigma$. According to the holographic proposal the
quantity \be S_{hol}={Area(\gamma)\over 4G_5}~~, \lb{shol} \ee
computed on the gravity side with Newton's constant $G_5$, is
identical to entanglement entropy computed for the quantum fields
defined on the $4d$-boundary of the space-time (\ref{metric}). The
correspondence is easily established for the UV divergent part of
the entropy as soon as we note that \be Area(\gamma)={L^3\over
2\epsilon^2}A(\Sigma)+..\ , \label{area}\ee where $..$ stands for
the subleading terms. Moreover, in the AdS/CFT correspondence one
has a relation \be {\pi L^3\over 2G_5}=N~~, \lb{N} \ee where $N$
is the number of fields (species) in the boundary theory. In fact,
the theory on the boundary is a $SU(n)$ Yang-Mills theory,
$n=\sqrt{N}$ is the number of colors. Thus, using the holographic
proposal one obtains  \be S_{hol}=N{ A(\Sigma)\over
4\pi\epsilon^2}~~, \lb{SN} \ee which is the right expression for
entanglement entropy of $N$ relativistic fields in four space-time
dimensions. (Note that the holographic UV regulator $\epsilon$ is
not identical to the proper time regulator $\epsilon$  appearing
in the heat kernel calculation.)

We now notice that the induced metric on a slice of  constant $t$
of (\ref{metric}) does not depend on the parameter $z$. Thus, for
a given surface $\Sigma$ on the regularized boundary the minimal
surface $\gamma$ and its area are the same for any value of $z$.
Assuming that the holographic description of entanglement entropy
can be extended to other values of $z$ we obtain that the entropy
then is still proportional to the area of $\Sigma$ and is UV
divergent in the same way as for the relativistic fields. For
$z=2$ the theory on the boundary is a non-relativistic theory
described by the Schr\"{o}dinger operator. The holographic
description thus is consistent with our direct calculation of
entanglement  entropy in a field theory with the Schr\"{o}dinger
operator.

Furthermore, from the fact that neither of the  quantities that appear
in (\ref{shol}) depends on $z$ it is natural to conclude that relation
(\ref{N}) still holds for value $z=2$ so that $N$ in this case is
the number of degrees of freedom in the non-relativistic theory.
The holographic formula (\ref{SN}) then is consistent with the
findings in this note: $N$ non-relativistic fields
 produce same entanglement entropy as $N$ relativistic fields. Clearly,
this statement and the equation (\ref{SN}), can be extended to any
dimension $d$. We note that in higher dimension $d$  entanglement
entropy  is given by expression (\ref{ent}) (the exact numerical
prefactor may depend on the regularization scheme) and is
invariant under the scale transformations (\ref{trans}). For $z={3\over 2}$
the holographic entropy was earlier calculated in \cite{DD} and agrees with our result.

We also note that the invariance under transformations
(\ref{trans}) with $\lambda=2$ helps to understand the structure
of the heat kernel (\ref{TrKD}).  Under (\ref{trans}) the proper
time $s$  rescales in the same manner as $r^2$, $s\rightarrow
\lambda^2 s$. Since the volume $V$ changes as $V\rightarrow
\lambda^{d+1}V$ one needs $s^{d+1\over 2}$ in the denominator of
the volume term in (\ref{TrKD}) for the invariance under
transformations (\ref{trans}).

\bigskip

\noindent{\it The interacting fields.} So far we have considered
the  non-relativistic fields without interaction. The holography
is however supposed to describe a strongly coupled field system.
In order to check the holographic predictions we have to introduce
interaction. The interaction can be included by adding a potential
term $\int d^dx V(\varphi)$ to the classical action, here
$\varphi$ is a set of fields in question.

In the  one-loop approximation one splits
$\varphi=\varphi_c+\varphi_q$, where $\varphi_c$ is the classical
background field and $\varphi_q$ is the quantum field. The
integration over $\varphi_q$ then  reduces to calculation of the
functional determinant of operator ${\cal D}_E+M^2(\varphi_c)$,
where $M^2\equiv V''(\varphi_c)$. If there are $N$ fields, for
simplicity we assume that $M^2$ is the same for all fields. A
generalization of this simplified situation is however
straightforward. The heat kernel of this operator is the product
$e^{-s{\cal D}_E}\cdot e^{-sM^2}$. Using the already calculated
trace (\ref{TrKD}) we obtain for the entropy of interacting fields
in $d$ dimensions \be S_{(d)}={N\over 12(4\pi)^{d-2\over
2}}A(\Sigma)M^{d-2}\Gamma(1-{d\over 2},M^2\epsilon^2)~~,
\label{Eint}\ee where we used that $\int_{\epsilon^2}^\infty
s^{-d/2}e^{-M^2s}=\Gamma(1-{d\over 2},M^2\epsilon^2)$. Clearly,
the leading UV divergence of the entropy (\ref{Eint}) is again
(multiplied by $N$)  (\ref{ent}) and is thus not affected by the
presence of the interaction in the action. This is consistent with
the holographic calculation.  As we have seen, the holography
predicts that the leading UV term in the entropy (\ref{SN}) is the
same as for free fields.

The interaction however shows up in the sub-leading UV divergent
and UV finite terms. For instance, in four dimensions we find \be
S={NA(\Sigma)\over 48\pi}  [{1\over
\epsilon^2}+M^2(\varphi_c)(\gamma-1+\ln(\epsilon
M(\varphi_c))^2)]\lb{ent4} \ee valid both in relativistic and
non-relativistic cases. In the holographic description the
subleading terms in the entropy  appear as the subleading  terms
in the area of minimal surface (\ref{area}). The   parameter $M$
in (\ref{ent4})  depends on the (constant) background field
$\varphi_c$ so that one may
 apply the  renormalization technique similar to one developed in \cite{DF}  for cosmic
 strings.
We also notice the earlier work on entanglement entropy in
relativistic $O(N)$ model \cite{ON}.

\bigskip

\noindent{\it Polynomial field operators, symmetry and the entropy.} The
symmetry argument helps to obtain the general structure of the
entropy in the case when the non-relativistic theory is described by
a polynomial field operator \be {\cal
D}_P=-\partial_\tau^2+m^{2(1-n)}(-\partial_i^2)^n ~~.\lb{DP} \ee
The heat operator $(\partial_s+{\cal D}_P)$  in this case is  invariant under the
transformations \be &&x_i\rightarrow \lambda x_i~,\ \ \tau\rightarrow
\lambda^n\tau~, \ \ s\rightarrow \lambda^{2n}  s~~,\nonumber \\
&&{\rm and}\ x_i\rightarrow \beta x_i~,\ \ m \rightarrow
\beta^{{n\over 1-n}}\ m~~. \lb{resc} \ee The trace of the heat
kernel then should have the form \be &&\Tr_\alpha e^{-s{\cal
D}_P}= a{m^{(d-1)(n-1)\over n} \over s^{d+n-1\over 2n}}\ \alpha V
\nonumber \\
&&+b C_2(\alpha)
({m^{n-1}\over s^{1/2}})^{d-2 \over n} A(\Sigma)~~, \lb{TrKG}\ee
which is invariant under (\ref{resc}), where $a$ and $b$ are some
constants which may depend on $n$ and $d$. Entanglement entropy in
a quantum field theory described by the polynomial  field operator
(\ref{DP}) is \be S\sim
({m^{n-1}\over \epsilon})^{d-2\over n} A(\Sigma)~~.\lb{Sd} \ee
For  $n>1$ the UV divergence of entanglement entropy becomes
milder than  in the relativistic case (\ref{S}).

\bigskip

\bigskip

\noindent{\it Conclusions.} Entanglement entropy is determined by
the short-distance correlations in the field system. That's why it
is surprising that the non-relativistic theory is characterized by
the same entropy as a relativistic field theory although the
short-distance behavior of two theories is different. (Indeed, in
d dimensions  the equal time correlation function is $1/|{\bf
x}|^{d-2}$ for a relativistic scalar field and $\delta({\bf x})$ for a
non-relativistic field.) In the present note we have demonstrated
this by a direct calculation of entanglement entropy using the
replica method and by extending the holographic description of
entanglement entropy to the non-relativistic theories which have dual
gravity descriptions. The agreement between two methods, in particular,  indicates
that the holographic description of entanglement entropy should have
wider applications.

The fact that we have the area law  in the non-relativistic case
may have some interesting consequences. In particular,  the area
law could be checked experimentally. One would have to prepare a
pure state in a condensed matter system available at the
laboratory and then arrange a situation when part of the system is
``hidden'' for observations. This may be yet another way to find
in condensed matter
 the features typical for black holes, the analog of
the Hawking radiation \cite{sonic} is the well known other
example.

\bigskip

This paper is based on the talk which the author gave at ``Sym\'etries
non relativistes : th\'eorie math\'ematique et applications
physiques'', Tours, 23-24 juin 2009. The helpful discussions with
A. Barvinsky, D. Fursaev, O. Lysovyi are kindly acknowledged.

\end{document}